\documentclass[12pt]{article}
\usepackage{graphicx}

\title{Slow crack propagation through a disordered medium:  critical transition
and dissipation}

\author{G. Pontuale$^{1}$, F. Colaiori$^{2,3}$ 
A. Petri$^{1,3}$\\
\small $^{1}$ CNR - Istituto   dei Sistemi Complessi, \\
\small via del Fosso del Cavaliere100, 00133 Roma  - Italy\\
 \small  $^{2}$ CNR - Istituto dei Sistemi Complessi, Dipartimento di Fisica,\\
\small Sapienza Universit\`a, P.le A. Moro 5, 00189 Roma - Italy\\
\small $^{3}$ Dipartimento di Fisica, Sapienza Universit\`a,\\ 
\small P.le A. Moro  5, 00189
Roma - Italy}

\begin{document}
\maketitle
\begin{abstract}We show that the intermittent and self-similar fluctuations displayed
by a slow crack during the propagation in a heterogeneous medium  can be
quantitatively described by an extension of a classical statistical  model for
fracture.  The model yields the correct dynamical and morphological scaling, and
allows to demonstrate that the scale invariance originates from the presence of
a 
non-equilibrium, reversible, critical transition which in the presence of
dissipation gives rise to  self organized critical
behaviour.
\end{abstract}

\section{Introduction}

The failure of materials is a complex and complicated process  exhibiting
broad  phenomenology. The fracture of heterogeneous media under slow external
loading
displays intermittent dynamics  and scale invariance,
features observed in different phenomena involving a huge range of length and
time scales, from nano-plasticity \cite{dimiduk06} and 
micro-fractures   \cite{chakrabarti97,petri94,garcimartin97,salminen02}
 to earthquakes \cite{turcotte92,corral04,kawamura12}.
In the  propagation of a slow crack, the front advances through  a movement that
is statistically
stationary but characterized by sudden and intermittent self-similar bursts, or
{\it
avalanches}. Bursts
area and duration are power law distributed, and the crack profile is
self-affine
(see {\it e.g.} \cite{bonamy09,bonamy10} and refs. therein). Besides having
practical relevance, this  phenomenology, known as  ``crackling  noise''
\cite{sethna01},
characterizes also the dynamics of other interfacial phenomena in disordered
media, like imbibition, wetting, friction, and hysteresis in ferromagnets.

In analogy with equilibrium phenomena, the absence of typical scales is
considered a mark of some underlying critical transition
\cite{dickman00}. This  conjecture has led  in the last decades to the
formulation  of simplified statistical models for fracture
\cite{chakrabarti97,roux90,alava06,pradhan10} with the aim of understanding the
origin of scale invariance, and identifying possible universalities 
in different systems.
However, it is diffuse opinion \cite{bonamy08,rosti10} that these models  have 
until now
failed in catching what are thought to be the right power laws characterizing
both the failure of
various materials  and the seismic behaviour accompanying earthquakes.
Extant models yield exponents significantly higher than the experimental
observations \cite{bonamy08}, and make the right values ``a target for
theoretical models'' \cite{alava06}.
Moreover, they fail in  reproducing the microscopic statistics in
quasi-stationary situations
\cite{alava06}, making the link with a reversible transition problematic.

On general grounds, observation of scale invariance in many natural
phenomena  has led to the concept of  Self Organized Criticality (SOC)
\cite{bak88,christensen05}, in which a system spontaneously sets close to a
critical
point, in contrast with ordinary critical phenomena where some parameter
needs to be finely adjusted.  Attempts have been made to relate the self-similar
fluctuations observed in fracture phenomena  with SOC dynamics
\cite{petri94,guarino98,caldarelli96}, and 
cellular automata have been devised with the aim of reproducing the power laws
observed  in
fractures and earthquakes \cite{kawamura12,christensen05,chen91}. SOC models
have been very helpful
and  intuitive in illustrating the general properties of the critical behaviour;
however
their matching with real systems is still difficult in relation to several
points, in particular 
the correspondence between model parameters and real ones
\cite{turcotte99}. So the  critical transition beneath the fracture
phenomenology has
remained elusive, even if more evidences have been recently emphasized
\cite{bonamy08}.

In this paper we show for the first time that the critical behaviour
characterizing the slow fracture and crack propagation in a heterogeneous medium
originates indeed from the existence of a non-equilibrium  critical phase
transition,
separating two distinct phases: an active phase, in which the fracture
propagates indefinitely,  and a dormant one, in which the system is
quiescent.
We shall introduce an extension of a widely employed statistical model,
described in the next section, and use it
to reproduce the  propagation of a planar crack. It will be seen in the
following sections  that this model yields  intermittent and self-similar
dynamics in quantitative agreement with the dynamical and morphological scaling
measured in recent experiments  \cite{maaloy06,bonamy08} and in simulations
\cite{bonamy09,laurson10}. Moreover, a same universal scaling is found
to characterize the avalanche area, stress, and energy. The nature of the
critical transition will be discussed in a separate section. Finally we will
show that dissipation  is crucial in
generating the observed intermittency by setting the systems at the edge of
criticality, and connecting the process to the SOC phenomenology.

\section{The model}
The model we adopt is an extension of the classical Fiber Bundle Model (FBM),
which in its original
formulation \cite{daniels45} consists of a set
of \textit{N} parallel fibers, each one having a random, quenched,
breaking strength drawn from some identical probability
distribution. Applying a slowly increasing load parallel
to the fibers axis,  more and more fibers break and redistribute their stress
among those still intact, which in turn can break, generating
avalanches in a domino effect. The  probability distribution for the number of
fibers broken in an avalanche is a
power law, with a cutoff at large values. The dynamics is non stationary, as
the number of intact fibers decreases with increasing load, and at a critical
stress
the system undergoes  global and irreversible failure.  While in its
original formulation FBM is a mean field  model \cite{sornette89,hansen92}, 
several variants have been also
devised to take into account dimensionality and different stress redistribution
rules among fibers
\cite{alava06,pradhan10,hansen94,virgilii07,dalton10,hidalgo02,lehmann10}.

In order to describe the propagation of a crack front during its
quasi--stationary
regime, we have complemented the classical FBM with two rules:
\textit{1) Fiber regeneration}: Each time a fiber beaks, it is  
replaced  by
another one, with new random breaking strength and zero initial stress.
\textit{2) Energy dissipation}: When a fiber breaks, part of its elastic energy
is dissipated. Thus only a fraction of its stress is transferred to other
fibers, while a complementary fraction is lost.  

These rules characterize the Dissipative Regenerating Fiber Bundle Model (DRFBM)
investigated in the present paper, 
and are motivated by the following considerations:
 \textit{1) Regeneration.}  The replacement of broken with intact fibers mimics
the crack
propagation, in which regions previously hedged from the stress get involved as
the
overloaded zones yield. The propagation of a planar crack, orthogonal to
the fibers,
can thus be described by a 1d version of the model. During each avalanche
broken fibers
are replaced by new ones, each failure event representing an onwards local
motion of the crack, as will be explained in detail in the section dedicated to
the roughness.

\textit{2) Dissipation.} The dissipation of elastic energy in real fractures may
happen
by several mechanisms: in the Griffith's description of a perfectly brittle
medium it is due to the cost of opening the crack. In the general case other
processes can contribute, such as the emission of elastic waves and the plastic
deformation. Whereas the introduction
of dissipation in the standard FBM  would just cause a slow down of the
dynamics,
and a delay of the final breakdown, we will show that in the DRFBM it is crucial
for the onset of intermittence in the motion of the crack front.

\begin{figure}
\begin{center}
\includegraphics[width=60mm,angle=-90]{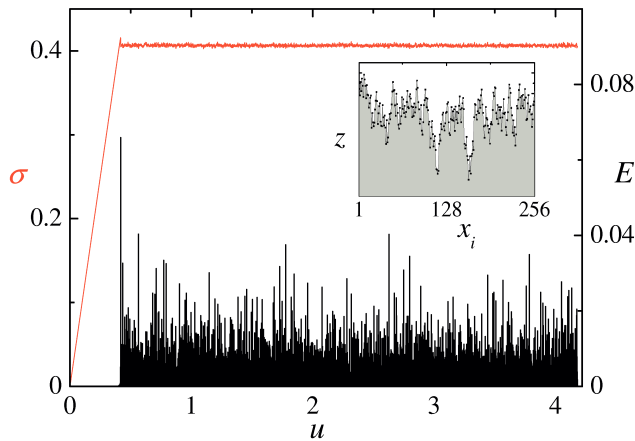}
\caption{\label{fig1} Avalanche energy (lower curve) and
instantaneous bundle stress (upper curve)
in the DRFBM as functions of the applied external strain,
 from numerical simulations with $N=10^4$ and $\delta=10^{-3}$.
Inset: a snapshot of the crack front obtained by the 1--d DRFBM, as described in
the section discussing  the roughness}.
\end{center}
\end{figure}

We consider here a bundle of harmonic fibers
 with identical unitary elastic modulus, each one having a random, quenched,
breaking threshold $t$
extracted from an identical probability density  $p(t)$.
The  dynamics of the DRFBM can be summarized as follows: starting from zero
initial stress,
all the fibers are subjected to a same slow increase of strain $u$,
until the weakest fiber breaks and it is replaced by an
intact fiber having zero stress and a new random threshold.  Soon after,
a fraction $(1-\delta)$ of the released stress
is  redistributed within the bundle according to some rule, while  the
remaining fraction $\delta$ is lost.
The redistributed  stress may cause the breaking of one or more other fibers,
which in turn will
redistribute part of their stress, and so on. Different fibers broken from a
same redistribution process are all regenerated at the same time. In the model
we assume
a separation between the timescales of the internal avalanches dynamics and the
driving field rate, keeping  the external strain fixed during avalanches. An
avalanche stops when the stress transferred from the broken fibers
causes no further failures.
At this point the bundle is subjected to a new strain increase until a
fiber breaks and so on, and the dynamics
evolves under such an adiabatic driving. It is worth noting that
each fiber bears a history dependent stress, resulting from summing, from the
epoch of its generation, the internally redistributed stress and the externally
applied strain.

\section{The propagation of a planar crack}
Here we consider the DRFBM model in $d=1$, with the fibers placed on a line and
subjected to periodic boundary conditions. Each site is identified by its
discrete
coordinate $x_i$, to which it is associated a set of successive fibers
$i_\alpha$. Each time a fiber $i_\alpha$ breaks it is replaced by a new one,
$i_{\alpha+1}$, and the crack front moves onwards. According to the rules that
will be spedified in the section devoted to the roughness, this gives rise to an
irregular crack front profile, like the one sketched in the inset of fig. 1.
Since only one fiber at time is present at a given site, we shall from now on
identify the current intact fiber with  only the site index $i$ and the related
stress with $s_i$, unless otherwise specified. 

In a crack front each point generates stress on
other points proportionally to the inverse  square of the relative distance 
\cite{GR89,tanguy98}.
Therefore we adopt a distance
dependent stress redistribution rule \cite{hidalgo02} such
that if the fiber at $x_i$  breaks under the stress $\tilde{s_i}$,
the stress of the fiber at $x_j$ increases by an amount
\begin{equation}
\label{elastrange}
\Delta s_j = \kappa \, \, \frac{\tilde{s_i}}{(x_i-x_j)^2} \, \, \,
\mbox{for} \, \, \, j \neq i \, \, \, \mbox{and} \, \, \, \Delta s_i=0, 
\end{equation}
where $x_i-x_j$ is taken modulo $N$, and
$\kappa$ is a $\delta$ dependent normalization factor assuring that
a fraction $\delta$ of $\tilde{s}_i$ is dissipated:
$\sum_{j} \Delta s_j = (1-\delta) \tilde{s}_i$.

Figure 1 shows the behaviour of  the 1--d DRFBM at increasing applied strain,
obtained  from the simulation
of a system with  $N=10^4$,  $\delta=10^{-3}$,  and random fiber thresholds $t$
extracted uniformly in
$[0,1)$.
The lower curve represents the energy of avalanches  $E= \sum_f
\tilde{s}^2_f/2$, where the sum is over
all the failures in the avalanche.   We observe that at low strain there are
only a few rare and small failure events,
but eventually a large avalanche starts, leading the systems to a state where
avalanches of any size occur.
At that stage each increase in stress due to the externally applied strain
balances in average the dissipated stress, and the total  bundle stress
$\sigma=\sum_i s_i/N$ reaches  a statistically stationary value $<\sigma>$, as
shown by the upper curve in the figure.

We have extensively investigated the 1--d DRFBM  with different values
of dissipation $\delta$ and size $N$.
For $\delta \neq0$,  under increasing strain
the system reaches a state
in which the bundle stress $\sigma$ is statistically stationary.
The case  $\delta = 0$ is
special  and will be discussed in a devoted section.
Besides the energy $E$, we have computed several quantities for each avalanche,
and their probability distributions : The  area $A$, defined as the total number
of failed fibers;
the  duration $T$, as the number of regeneration processes
occurring within the avalanche, counting as sigle process the simultaneous
regeneration of more fibers; the stress $S$, as the sum of the stress
born by all the fibers broken in the avalanche.
In addition,  since isolated avalanche clusters
can take place at the same time in different regions of the crack front
\cite{bonamy08,maaloy06,laurson10},
we have also considered the statistics of the area $C$
of isolated clusters,  {\it i.e.} of
locally connected but mutually disconnected failures in the same avalanche.

It is seen that  the bursts display self-similar features in all the quantities
considered.
For large enough avalanches the distribution for each quantity,
$y$, is well described by
\begin{equation}
\label{ffs}
p(y) \simeq  y^{-\tau_y}f_y(y/y_{o}).
\end{equation}
The exponents  $\tau_y$ do not depend
on the dissipation rate $\delta$, which instead determines the
 cut--off values  $y_{o}$.
 By assuming a power law dependence,  $y_{o} \simeq \delta^{-\psi_y}$, the
distributions  of a same quantity for different $\delta$ can be collapsed onto a
unique curve.
Figure 2 shows, on double logarithmic scale, the collapsed probability
distributions of (a) the
cluster avalanche area $C$, (b) the  total avalanche area $A$,
(c) the  avalanche duration $T$, as  obtained from numerical simulations of
the  1--d DRFBM with $N=10^4$ and for different values of $\delta$.

By means of the data collapse we have got as  best estimates for the exponents:
 $\tau_A = 1.2$ and $\tau_C = 1.5$ for the total and cluster area, respectively,
and $\tau_T = 1.5$ for the duration.
These values, resumed in the table I, strictly agree with those
found in the experiments \cite{bonamy09,maaloy06}, in linear fracture models
 \cite{bonamy08} with following rediscussions
\cite{bonamy09,bonamy10,laurson10}, and in depinning models (see {\it e.g.}
\cite{bonamy08} and refs. therein). 
The  rescaling  cut--off  exponents $\psi_y$ are also reported in the table I.
Since it is expected that the average avalanche obeys the scaling  $<y>
\approx \delta^{-\phi_y}$,
the exponents must be  consistent with  the relation  $\phi_y=\psi_y(2-\tau_y)$,
that is indeed the case.

\begin{table}
\begin{center}
\caption{\label{table1} Dynamical scaling exponents for a planar crack from the
1--d DRFBM}
%\begin{ruledtabular}
\begin{tabular}{cccc}
& $\tau$ &  $\psi$ & $\phi$ \\
\hline
C & 1.5 & 1.4 & 0.7 \\
A & 1.2 & 1.3 & 1.0 \\
T & 1.5 & 0.7 & 0.4
\end{tabular}
\end{center}
%\end{ruledtabular}
\end{table}

We have also computed the distribution of the energy $E$ and stress $S$, for
both total and cluster avalanches, and  we have found that they are described by
the same exponents of the respective avalanche areas 
$\tau_C$ and $\tau_A$,  indicating that the scaling of the size fluctuations is
a universal feature, irrespective of the measured quantity.

Finally we have checked that all results do not depend on the
probability distribution $p(t)$ from which the fiber threshold strenghts are
extracted, provided
 that this vanishes rapidly enough as $t \to \infty$.
Finite size effects  are not generally displayed
as they appear only for very small values of $N \cdot \delta$.

\begin{figure}
\begin{center}
\includegraphics[width=60mm,angle=-90]{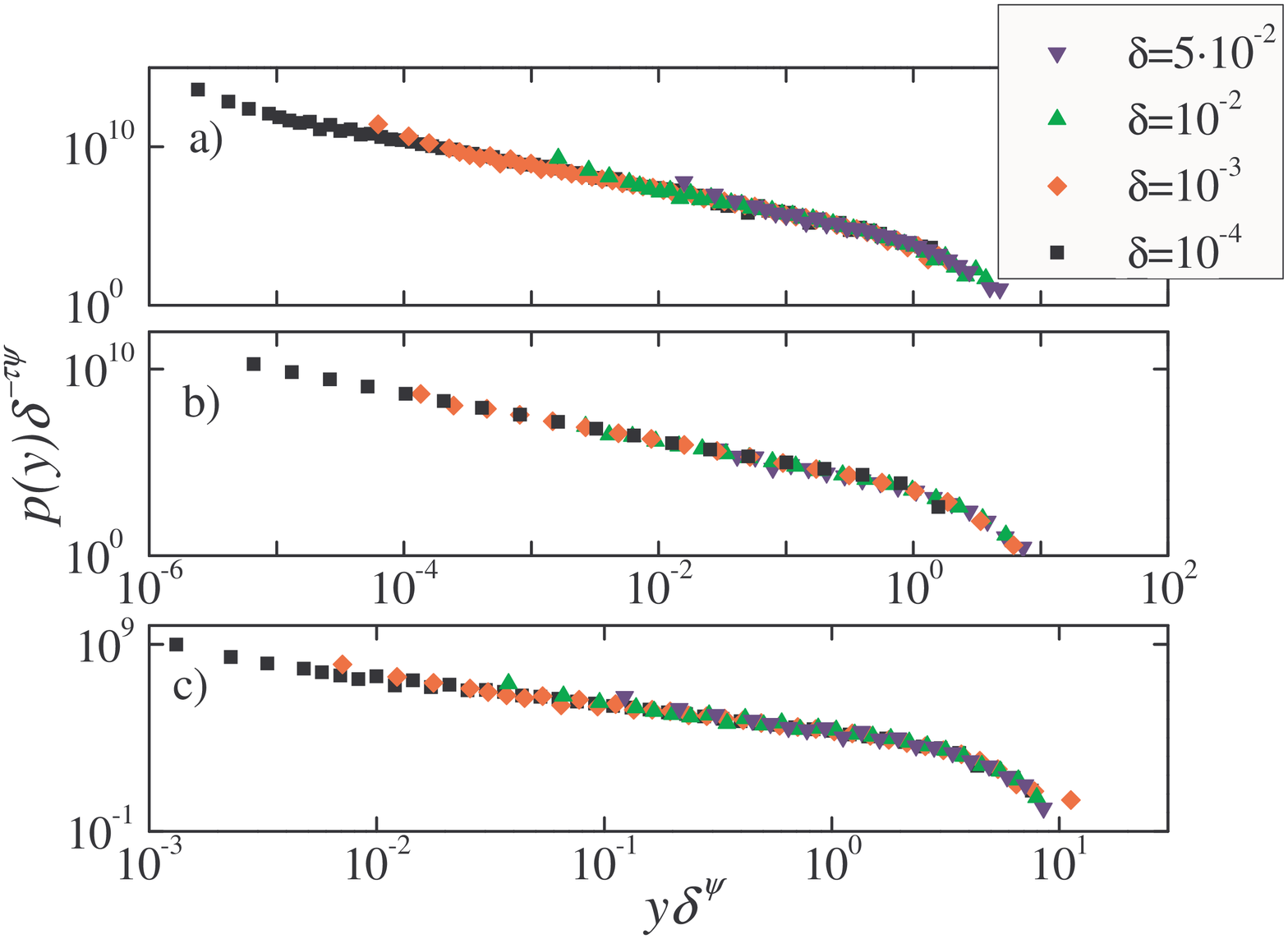}
 \caption{\label{fig2}
Collapsed probability  distributions for the avalanches in the planar crack
front dynamics as resulting
from the DRFBM, relative to: (a) cluster avalanche area $C$; (b) total  area
$A$;
 (c) duration $T$. Relevant exponents are reported in table 1.}
\end{center}
\end{figure}

\section{The roughness of the front}

The self-similar features of crack profiles have been subject of study since
when they were
first observed  in the surface of fractured metals  \cite{mandelbrot84}.
A wide amount of work has then been devoted to characterizing different
morphological scalings,  clarifying their origin  and looking for classes of
universality \cite{bonamy10,bouchaud97}.
It also required time to explain some apparent contradictions between different
experiments, and discrepancies  bewtween some experimental results and models
for elastic depinning 
(see {\it e.g.} \cite{dalmas08,bonamy10}), but it seems now rather well
established that different scaling regimes can be attributed to the interplay between
elasticity,
disorder and possible spatial correlations in the disorder strength
\cite{laurson10b,santucci10}. 

In order to define the position of the crack front in the 1--d DRFBM, we
assume that
each time a fiber yields, the crack advances locally -- perpendicularly to the
fibers and 
to the {$x$} axis -- a distance proportional to the stress $\tilde{s}$ stored in
the broken fiber.   
Since fibers are harmonic, the total perpendicular displacement at
$x_i$ is given by
 $z_i=\sum_{\alpha} \tilde{s}_{\alpha_i}$  where
$\alpha_i$ counts all the consecutive failures occurred at site $i$ up to the
current epoch.

\begin{figure}[h]
\begin{center}
\includegraphics[width=60mm,angle=-90]{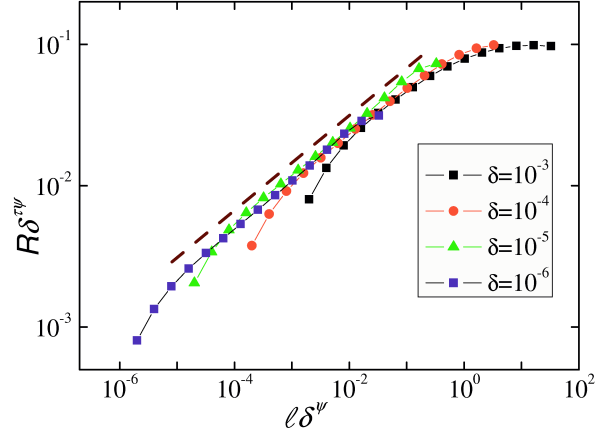}
  \caption{\label{fig3} Scaling of the crack front roughness $R$ in the 1--d
DRFBM. Curves from systems with different dissipation have been collapsed by
assuming the same dependence of 
the upper cut-off on $\delta$. The dashed line has slope $\zeta=0.35$.}
\end{center}
 \end{figure}

The scaling properties of the roughness can be evaluated through different
quantities.
Here we compute 
\begin{equation}
\label{rough}
R^2_\ell =  \langle (z_{i+\ell} - z_i)^2 \rangle
\end{equation}
that is expected to scale as $\ell^{\, 2\zeta}$. For the 1--d DRFBM we find a non
trivial power law behaviour of $R_\ell$.  Figure 3 shows  the values of $R_\ell$
as function of $\ell$ for
uniform initial distribution of the fiber strength,  $N=2^{16}$ and different
dissipations $\delta$. The crack profile is self-similar in a wide range and
it can be noticed that even in this case the dissipation plays the fundamental
role of setting the upper cut-off $\ell_{o}$ of the scaling range. Different
curves
corresponding to different values of dissipation can in fact be collapsed
assuming
the  scaling  (\ref{ffs}) with a dependence
of  the form $\ell_{o} \simeq \delta^{-\psi_{\ell}}$. A best fit yields
 $\zeta = 0.35$ in agreement
with  those  found in experiments and simulations for uncorrelated and not too
strong disorder \cite{santucci10,laurson10b}. 
We also find $\psi_{\ell} = 1.00$.

Very recently, the transition betweeen two different scaling regimes has  
been also observed in a different FBM based model  \cite{gjerden12} in which  the  fibers
interact through a 2--d  kernel. The model  shows that, as the  Young 
modulus increases, the front becomes rougher and overhangs and damage 
islands appear, the  exponent in the soft elasticity regime being in  agreement with  
that observed in the DRFBM.

\section{The critical transition and the  effect of the dissipation}

As discussed in the introduction, a major point is to establish whether
a critical transition is at the origin of the scale free fluctuations observed
in the propagation of the crack front.
In fact, linking irreversible intermittent phenomena to a reversible
transition may be of help in the identification of different universality
classes and of key working mechanisms.\\

In a phase transition, a system must exist in at least two different states
separated by a critical point. In order to address this issue we have
investigated the system behaviour with null dissipation ($\delta=0$). 
Note that in this case all the energy fed remains in the system, and the bundle
stress $\sigma=\sum_i s_i$ equals the totally applied strain.

As usual,  an external strain was supplied in a quasi-static mode, starting with
zero initial stress.
The $\delta=0$ case displayed initially a transient phase similar to the case
$\delta \neq 0$.  As the bundle stress was increased, larger and larger
avalanches appeared, until eventually the bundle stress $\sigma$ reached a
maximum (like in fig.1 for $\delta \neq 0$) and a very large avalanche was
triggered.
While in the case  $\delta \neq0$ this peak is necessarily followed by a decay 
and  $\sigma$  sets at a lower stationary value, in the $\delta=0$ case  the
domino effect provokes a neverending avalanche, and the system remains
indefinitely in the active state, the redistributed stress bouncing among the
fibers.
Hence,  for null dissipation, the system is driven to a critical point  where
the crack propagation sustains indefinitely, whilst in the presence of
dissipation, however small, 
it goes back below the critical point, and only new a stress increase will
trigger more avalanches.

We have simulated the 1--d DRFBM with no dissipation and different sizes, and
computed the
fraction of active fibers $\rho$, {\it i.e.} the average number of fibers
simultaneously breaking during avalanches, at varying bundle stress. The results
are shown in fig. 4, where it is seen that close to a critical value $\sigma_c$
the activity suddenly departs from zero and monotonically increases. We have
also computed the variance of
$\rho$, that is proportional to the system susceptibility. It is seen to display
a peak close to  $\sigma_c$, and to increase linearly with the system size, as
expected from finite size scaling in
d=1.

These features are the hallmark of a critical transition and show that 
for $\delta \neq0$ the system is at the lower edge of criticality. In fact, in
this case, the peak stress attained  before the stationary state, visible in
fig.1., corresponds to the critical value 
$\sigma_c$. 

Finite size scaling study and detailed analysis of this transition will be
reported
elsewhere. We stress here that the resulting picture is the one described by
Dickman {\it et al.} \cite{dickman00} for the onset of Self Organized
Criticality (SOC), where the system oscillates continuously between an active
phase, to which it is driven by the external field, and an absorbing phase where
it falls because of the dissipation. This demonstates in a clear and definite
way that  crack propagation is a realization of SOC, as also recently
conjectured \cite{bonamy08}.

Dissipation, that has been treated as an irrelevant parameter in linear elastic
fracture models \cite{tanguy98,bonamy08,laurson10} and  useful regolarizing
factor in depinning models \cite{ledoussal09},
turns out to be the key factor for the onset of intermittence, establishing a
clear link between  fracture processes and  SOC.  In the DRFBM it occurs in the
bulk, at variance with usual 
Self-Organized-Critical (SOC) 
models ({\it e.g.} sand piles) where it is limited to the boundaries in order to
obtain the cut-off to disappear  when taking the thermodynamic limit. As a
matter of fact, internal dissipation is hardly unavoidable in real macroscopic
systems, and  there is growing evidence that it governs real manifestations of
SOC  \cite{bonamy08,colaiori08,baldassarri06,peters06}. It does not affect the
exponents and may well be at the origin of the large scale cut-off always
present in real systems, in alternative, or in addition, to finite size effects.
Finally, it is also noticeable that, to our knowledge, a diverging
susceptibility had never been observed in SOC models, whereas it characterizes
natural phenomena \cite{peters06}.

\begin{figure}
\begin{center}
\includegraphics[width=60mm,angle=-90]{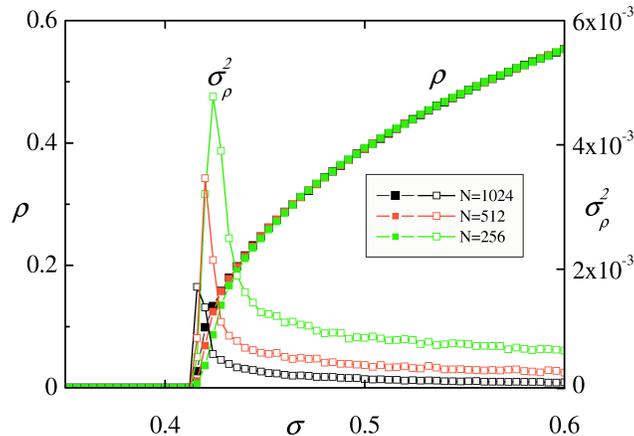}
  \caption{\label{fig4} Evidences of the critical phase transition displayed by
the 1--d DRFBM in the absence of dissipation. The fraction of active fibers
$\rho$ suddenly departs from zero above a
critical value of the bundle stress $\sigma$. The variance $\sigma^2_\rho$ shows
a peak that increases with the system size.}
\end{center}
 \end{figure}

\section{Discussion and summary}

We have described the propagation of a slow crack front in a heterogeneous
materials by means of a   stationary Fiber Bundle Model which includes internal
dissipation and fiber regeneration (DRFBM).
In the absence of dissipation the system can be in either of two distinct
phases, one active and the other quiescent, separated by a critical value of the
system stress at which the susceptibility diverges. The presence of dissipation,
 however small, drives an active system to a slightly subcritical state whereas,
by supplying stress, the system can be driven from the quiescent to the active
phase.  The  alternance of these two processes leads the system to a
statistically stationary state where the dynamics is scale invariant and
intermittent.

The 1--d DRFBM with the stress redistribution rule eq. (1) yields the correct
experimental scaling of both dynamical and morphological fluctuations of the
planar crack.  Universal statistics characterizes the burst area, energy and
stress, whereas dissipation sets the upper cutoff to the self-similar range.

The DRFBM establishes a link between SOC models and real systems and can be
easily extended to describe systems in higher dimension, like non planar cracks
and earthquakes, or can be employed for implementing preexisiting models
\cite{klein11}.
 It also presents interesting analogies with other breaking-healing contact
models, recently introduced to describe tribological  experiments, earthquakes
\cite{urbakh09} and biological systems  \cite{filippov11}, that would deserve to
be investigated.

\bibliographystyle{abbrv}
\bibliography{FBM-refs,sheargrbibv2,friction,crit}

\end{document}